\documentclass[aps,prb,preprint,superscriptaddress,showpacs,floatfix]{revtex4}
\usepackage{graphicx}
\begin{document}
\title{Lattice gas models of coherent strained epitaxy}
\author{V. I. Tokar}
\affiliation{
IPCMS-GEMM, UMR 7504 CNRS, 23, rue du Loess,
F-67037 Strasbourg Cedex, France
}
\affiliation{Institute of Magnetism, National Academy of Sciences,
36-b Vernadsky str., 03142 Kiev-142, Ukraine}
\author{H. Dreyss\'e}
\affiliation{
IPCMS-GEMM, UMR 7504 CNRS, 23, rue du Loess,
F-67037 Strasbourg Cedex, France
}
\date{\today}
\begin{abstract}
The harmonic Frenkel-Kontorova model is used to illustrate with an
exactly solvable example a general technique of mapping a coherently
strained epitaxial system with continuous atomic displacements onto a
lattice gas model (LGM) with only discrete variables. The misfit strain
of the original model is transformed into cluster interatomic
interactions of the LGM. The clusters are contiguous atomic chains of
all lengths but the interaction strength for long chains is
exponentially small. This makes possible the application of efficient
Monte Carlo techniques developed for discrete variables both in kinetic
and equilibrium simulations.  The formalism developed can be applied to
1D as well as to 2D systems.  As an illustrative example we consider
the problem of self-organization of 1D size calibrated clusters on the
steps of the vicinal surfaces.
\end{abstract}
\maketitle
\section{Introduction}
The phenomena of self-assembly and self-organization of coherent
(i.~e., dislocation-free) size calibrated nano- and atomic-scale
structures observed during the heteroepitaxial growth in some
systems~\cite{discovery,dots} are considered to be promising tools for
fabrication of microelectronic devices~\cite{devices}.

A major factor influencing the phenomenon of self-assembly is the
lattice size misfit (LSM) between the substrate and the growing
overlayer commonly encountered in heteroepitaxial
systems~\cite{mismatch}.  The LSM-induced strain is believed to be the
driving force behind the above phenomena \cite{mismatch,lannoo}. So an
adequate account of the strain effects should lie at the basis of any
theory of strained epitaxy.  Because strained systems exhibit
complicated kinetics and morphologies, analytic approach is difficult,
so a major technique in theoretical studies of strained epitaxy is the
kinetic Monte Carlo simulation.  The application of this technique,
however, is severely hampered by the necessity to simulate the
continuous atomic displacements. Because of this, atomistic models in
such simulations are currently restricted to rather small systems
consisting of only a few thousand atoms \cite{2D3D} while
experimentally observed 3D quantum dots sometimes consist of several
tens of thousand atoms each \cite{dots}.

Our research is based on the observation that as long as we are
interested only in the {\em coherent} structures, there is a
possibility to map the system onto a purely lattice model because in
the absence of dislocations the relaxed atom can only either deviate
from its symmetric position inside the same cell or to be displaced to
another cell but there always exists a lattice site of a regular
lattice to which this atom can be ascribed.  So our first goal is to
develop a formalism which would allow to map a coherent heteroepitaxial
system with the continuous variables onto a lattice gas model with only
discrete variables.

One of important goals of the heteroepitaxial studies is the
development of techniques of growth of 1D quantum wires (QW) which can
be used, e. g.,  for experimental investigation of the Luttinger model
of interacting 1D electrons \cite{nat_chains}.  Furthermore, the QW may
find application in microelectronics circuitry \cite{devices}, as well
as in magnetic memory devices \cite{magnetism}. The latter applications
would require the 1D structures of finite length. In the case of the
memory devices it is also desirable that these atomic clusters (or
chains) were of similar length and that they were arranged into
periodic arrays in order to simplify the memory access.  All these
requirements can be satisfied by self-organized size calibrated
structures similar to quantum dots of the 2D epitaxy \cite{dots}. A
phenomenological theory explaining the mechanism of formation of
quantum dots was proposed in Ref.\ \cite{lannoo}.  The theory is quite
general and can be applied to objects in any number of dimensions.  So
to illustrate the techniques developed in the present paper we will
study the conditions of formation of the 1D size calibrated monatomic
chains.
\section{The model} 
A simple approach to theoretical description of the misfit is provided
by the 1D Frenkel-Kontorova model (FKM) which is frequently been used
in qualitative~\cite{2d1d,model2} and semi-quantitative
studies~\cite{baskiPRL} of strained epitaxy.  To illustrate the
generality of our approach we first present the formalism appropriate
to the 2D epitaxy but in concrete examples we will restrict ourselves
to the 1D case.

We consider an ensemble of a fixed number ($N$) of atoms coherently
deposited on a surface with a square lattice of deposition sites. The
rectangular lattice geometry was chosen because it allows for the
separation of $x$ and $y$ variables (see Ref.\ \cite{2d1d} and below).
Let us first consider more general model with the energy of the
epilayer of the form
\[
U = \sum_{\bf i}n_{\bf i}V_s({\bf R_i + u_i}) 
+ \frac{1}{2}\sum_{\bf ij}n_{\bf i}n_{\bf j}V({\bf u_i 
+  R_i -  u_j -  R_j}),
\]
where ${\bf R_i}=(a^xi_x,a^yi_y)$ ($a^\gamma$ the lattice constants in
the two directions, $i_x$ and $i_y$ integers ), $n_{\bf i} = 0, 1$ is
the occupation number of site ${\bf i}$, ${\bf u_i}$ the atomic
displacement, $V_s$ the potential of interaction with the substrate,
and $V$ the interatomic potential.

The analysis of the system considerably simplifies at low temperatures
where it can be approximately reduced to a lattice gas model. This is
achieved by exploiting the fact that the residence time of atoms at the
lattice sites can be arbitrarily large due to the Arrhenius law obeyed
by the probability of activated hopping over the potential barriers
separating neighboring sites \cite{thereference}. The dynamics of the
variables ${\bf u_i}$, on the other hand, do not have any energy
barriers.  So at sufficiently low temperature these variables are
capable of reaching their thermal equilibrium distribution during the
time intervals between the atomic hops, i. e., with the atomic
configuration remaining unchanged.  Averaging over ${\bf u_i}$ will
leave us with an effective non-equilibrium free energy function
$F_{\text{eff}}$ of variables $n_{\bf i}$ only:
\begin{equation}
\label{F}
\exp(- F_{\text{eff}}/k_BT) = \int \prod_{\{n_{\bf i} = 1\}} d{\bf
u_i}\exp(-U/k_BT).
\end{equation}
This purely lattice model can be further used in both equilibrium and
kinetic studies.

For the purposes of qualitative analysis it will suffice to average out
the displacement variables in Eq.\ (\ref{F}) in the harmonic
approximation, i. e., by expanding $V_s$ and $V_p$ in the above
equation up to the second order in the displacement variables $\bf u_i$
\cite{baskiPRL,model3} which is valid for small $|{
u^\gamma_i}|/a^\gamma$.  This allows to perform the calculation exactly
(see below). However, the integration can be extended to approximately
account also for anharmonic terms. This extension, besides making the
approach more accurate, can also account for some qualitative phenomena
\cite{model2}.

To facilitate comparison with other studies based on the
Frenkel-Kontorova model \cite{2d1d,model1,model2,baskiPRL} we write the
harmonic approximation as the second order power series expansion in
$\bf u_i$ for $V_s$ and in $\bf u_i - u_{i-\gamma}$ for the pair
potential $V_p$:
\begin{eqnarray} 
\label{U} 
U &\approx& V_sN+\frac{1}{2}\sum_{{\bf i}\gamma}n_{\bf i}
k^\gamma_s(u^\gamma_{\bf i})^2 
+ \frac{1}{2}\sum_{\bf ij}n_{\bf i}n_{\bf j}
V_{\bf ij} \nonumber\\&&+ \frac{1}{2}
\sum_{\bf i\gamma}k^\gamma_p[(u_{\bf i}^{\gamma} 
-  u_{\bf i+\gamma}^{\gamma} -  f_\gamma)^2- f_\gamma^2]
n_{\bf i}n_{\bf i+\gamma}\nonumber\\
&=&V_sN + \frac{1}{2}\sum_{\bf ij}V_{\bf ij}n_{\bf i}n_{\bf j}
\nonumber\\&&+ \frac{1}{2}\sum_{\bf ij\gamma}k^\gamma_pD_{\bf 
ij}^\gamma u_{\bf i}^\gamma u_{\bf j}^\gamma  
+\sum_{\bf i\gamma}k^\gamma_pf_\gamma u_{\bf i}^{\gamma}
\nabla^{\bf \gamma}_{\bf i}n_{\bf i}n_{\bf i + \gamma}
\end{eqnarray}
where $V_s$ and $V_{\bf ij}$ are the values of the substrate and the
pair potentials at the lattice sites for the atom in the symmetric
position (the zero order approximation), $\gamma$ denotes $x$ or $y$
component; $f_\gamma=-V^\prime_\gamma/V^{\prime\prime}_{\gamma\gamma}$
(where $V$ is the pair potential) is interpreted as the misfit
parameter in the direction $\gamma$, $k^\gamma_s$ and $k^\gamma_p$ are
the second derivatives of the corresponding potentials, $D_{\bf
ij}^\gamma$ is the dimensionless dynamical matrix defined by this
equality, and $\nabla^{\bf \gamma}_{\bf i}F_{\bf i}\equiv F_{\bf
i}-F_{\bf i-\gamma}$. A major simplification is achieved under the
nearest neighbor (NN) approximation for the relaxation because, as is
seen from the last line of Eq.\ (\ref{U}), in this case the variables
$u_{\bf i}^x$ and $u_{\bf i}^y$ separate, and the relaxations along the
two directions are independent.

With approximation (\ref{U}) the statistical average amounts to the
Gaussian integration to give
\begin{eqnarray}
\label{F2}
F_{\text{eff}} =&& \bar{V}_sN+\frac{1}{2}\sum_{\bf 
ij}n_{\bf i}n_{\bf j}V_{\bf ij}
-\frac{k_BT}{2}\sum_\gamma\ln\det G^\gamma \nonumber\\ 
&&- \frac{k_pf^2}{2}\sum_{\bf ij\gamma}
G^{\gamma}_{\bf ij}\nabla^{\bf \gamma}_{\bf i}n_{\bf i}n_{\bf i + \gamma}
\nabla^{\bf \gamma}_{\bf j}n_{\bf j}n_{\bf j + \gamma},
\end{eqnarray}
where $G^{\gamma} = 1/D^\gamma$ and in $\bar{V}_s$ we gathered all
terms which are proportional to the total particle number, such as the
normalization of the determinant coming from the in-plane Gaussian
integration and a similar term from the integration along
$z$-direction.  We do not discuss these contributions here because in
the present study we restrict ourselves to systems with a fixed number
of deposited atoms $N$.  In case of necessity these terms can be easily
recovered. We also note the entropic contribution (the second term in
the first line) which naturally appears in our formalism and which was
shown to be crucial for proper description of the processes of
deposition~\cite{einstein1} as well as for the correct prediction of
the shape of the atomic clusters~\cite{einstein2}.

Because of the gradient factors, only the ends of contiguous chains of
atoms contribute into the last term of Eq.\ (\ref{F2}) leaving two
matrix elements of $G^{\gamma}_{\bf ij}$ in the sum over ${\bf ij}$
inside every chain: the diagonal one which we denote as $G^{(0)}_{l}$
($l$ the length of the chain; we omit the superscript $\gamma$ to
simplify notation) and the matrix element $G^{(l-1)}_{l}$ connecting
the two ends of the chain. Furthermore, in the NN approximation the
matrix $D$ is block-diagonal because the atoms belonging to different
chains do not couple. Therefore, the determinant factorizes and the
relaxation part of the free energy which consists of the terms in Eq.\
(\ref{F2}) containing $G$ takes the form
\begin{eqnarray}
\label{Frelax}
F_{\text{relax}}&=& -\sum_{\text{chains}} \{\frac{k_BT}{2}\ln\det 
\frac{G_l}{G_1}+k_pf^2[G^{(0)}_{l}-G^{(l-1)}_{l}]\}\nonumber\\
&\equiv&\sum_{\text{chains}}(-TS_l+W_l)\equiv\sum_{\text{chains}}E_l,
\end{eqnarray}
where the summation is over the chains consisting of two or more
atoms.  Besides, in the first term on the right hand side (r. h. s.) we
subtracted the single atom entropy term by assuming that it is
accounted for in $\bar{V}_s$ which we discarded to simplify notation.
It is easy to show that $F_{\text{relax}}$ can be expanded into an
infinite sum of multiatom interactions of the form $V_ln_{\bf i}n_{\bf
i+\hat{\gamma}}\dots n_{{\bf i} + (l-1)\hat{\gamma}}$, where
$\hat{\gamma}$ is the unit vector in the direction $\hat{\gamma}$. The
expansion coefficients are given by the expression
\begin{equation}
\label{V_l}
V_l=E_l-2E_{l-1}+E_{l-2}
\end{equation}
valid for all $l\geq2$ with $E_0=E_1=0$.
\section{The relaxation free energy}
\begin{figure}
\includegraphics[viewport =  70 70 275 415, scale = 0.7]{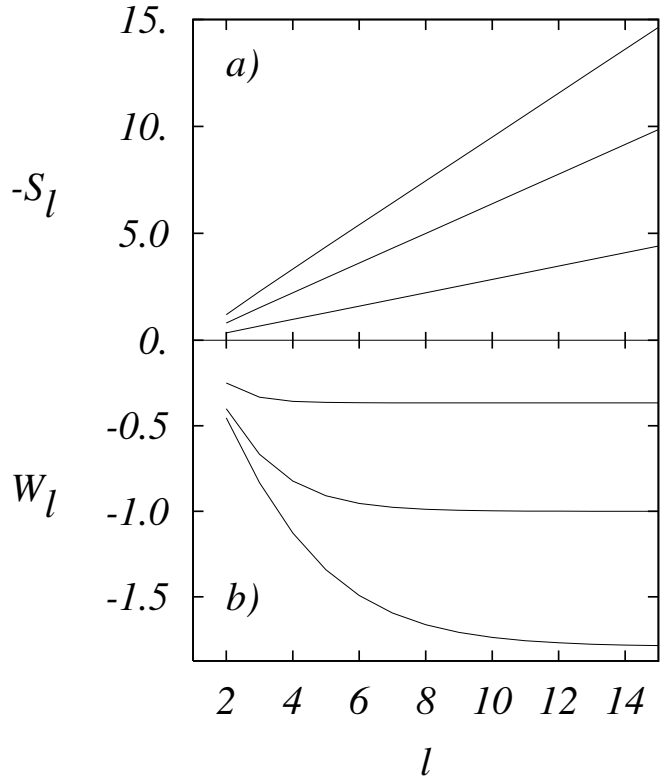} 
\caption{Relaxation entropy in units of $k_B$ (a) and relaxation energy
in units of $k_pf^2$ (b) for chains of length $l$. At both figures the
first curve from the horizontal zero axis corresponds to $\alpha=2$,
the second to $\alpha=0.5$, and the third one to $\alpha=0.2$ (note
difference in scale).}
\label{fig1}
\end{figure}

Omitting the superscript $\gamma$ to simplify notation, the matrix
$D_l$ for chain of length $l$ is defined according to Eq.\ (\ref{U}) as
\begin{equation} 
k_p\sum_{ij}(D_l)_{ij}u_iu_j=k_su_1^2+k_p(u_1-u_2)^2+k_su_2^2+\dots
+k_p(u_{l-1}-u_l)^2+k_su_l^2.
\end{equation} 
Hence, the square matrix $k_pD_l$ have the following structure:
\begin{equation} 
\label{kD}
k_pD_l=\left(\underbrace{
\begin{array}{ccccccc} 
 k_p+k_s &   -k_p   &  0   &\dots&0&0&0\\
   -k_p    & 2k_p+k_s&  -k_p   &0&\dots&0&0\\
   0    &   -k_p   &  2k_p+k_s&-k_p&0&\dots&0\\
\vdots &\vdots  &\ddots &  \ddots &\ddots&\vdots&\vdots\\
0& \dots &0&  -k_p  & 2k_p+k_s &   -k_p    &   0   \\
0&0&\dots   &0     & -k_p   & 2k_p+k_s &   -k_p \\
0&0&0&\dots  &      0   &   -k_p    & k_p+k_s
\end{array}}_{\textstyle l}
\right)
\end{equation} 
In Appendix \ref{appendix} we have shown that the matrix elements of
$G_l=1/D_l$ entering the expression for the relaxation free energy
(\ref{Frelax}) can be calculated with the use of recursion formulas for
the tridiagonal matrices as [see Eqs.\ (\ref{ GG}) and (\ref{
recursion})]
\begin{equation} 
\label{ GG1}
\left\{
\begin{array}{lcl}
G^{(0)}_{l+1} &=& 1/(1+\alpha-d_l)\nonumber\\
G^{(l)}_{l+1} &=& \det G_{l+1} =G^{(0)}_{l+1}b_l,\nonumber
\end{array}
\right.
\end{equation} 
where $\alpha = k_s/k_p$; $d_l$ and $b_l$ are generated by the
recursion relations
\begin{equation} 
\label{ recursion1}
d_{l+1} = 1/(2+\alpha - d_l)\quad\mbox{and}\quad  
b_{l+1} = d_{l+1}b_l.
\end{equation} 
The recursion relations (\ref{ recursion1}) at large $l$ drive $d_l$
and $b_l$ to the fixed point $d_\star, b_\star$:
\begin{eqnarray} 
d_\star &=& 1 + \frac{\alpha}{2} -\sqrt{\alpha + \frac{\alpha^2}{4}}<1\\
b_\star &=&0.
\end{eqnarray} 
This means that at large $l$ $G^{(0)}_{l}$ saturates to a constant
value while $G^{(l)}_{l}=\det G_{l}\sim d_\star^l$ [see Eqs.\ (\ref{
GG1})].  This in particular means that the entropic contribution into
$F_{relax}$ which is proportional to $\ln (\det G_{l})$ is
asymptotically linear in $l$ (see Fig.\ \ref{fig1}).

Two contributions into $F_{\text{relax}}$ calculated with these
formulas are shown in Fig. \ref{fig1}. The entropic contribution is
negative because in our canonical ensemble formalism we discarded the
entropy corresponding to $N$ isolated atoms which is reflected in the
denominator $G_1$ in the expression (\ref{Frelax}) for the free
energy.  This means, in particular, that when $k_p\to0$ the atomic
displacements within the cell became mutually independent (the on-site
pair interactions do not depend on the atomic positions inside the
cell), so $G_l\to G_1\hat{I}$, where $\hat{I}$ is the unit matrix, and
the entropy term tends to zero [see Eq.\ (\ref{Frelax}) and
Fig.\ \ref{fig1}]. When $k_p$ is not equal to zero, the atomic
displacements became restricted by its neighbors, so their entropy
diminishes. For large chains the interior atoms are practically in
translationally invariant environment, so for long chains the entropy
loss is proportional to $l$. The above restrictions on the atomic
displacements are the more stringent the larger $k_p$. This is
reflected in the slope of $S_l$ which is larger for smaller
$\alpha=k_s/k_p$ (see Fig.\ \ref{fig1}).

Because the entropic contribution in Eq.\ (\ref{Frelax}) is multiplied
by $T$, at low temperature it can be neglected.  The remaining term
$W_l$ is negative and saturates to a finite value at large $l$ (see
Fig.  \ref{fig1}b).  From Eq.\ (\ref{U}) it follows that the NN
interatomic interaction $V_{\text{NN}} = V_{\text{NN}}^{\text{ch}}
+k_pf^2$, where the first term ($V_{\text{NN}}^{\text{ch}}$) is some
``chemical'' interaction and the second term is the unrelaxed repulsion
due to LSM.  Thus, the relaxation energy in Eq.\ (\ref{Frelax}) is the
difference between the relaxed elastic energy of the chain
$E^{el}_{rel}(l)$ and the unrelaxed one $E^{el}_0(l)=(l-1)k_pf^2$ with
both energies being positive, as it should be for elastic energies.
Thus,
\[
W_l=E^{el}_{rel}(l)-E^{el}_0(l)
\]
and is negative because relaxation lowers the energy.  Now, if
$V_{\text{NN}}^{\text{ch}}$ is negative but such that $V_{\text{NN}}$
is not too large, then the reduced chain energy $[W_{l} +
V_{\text{NN}}(l-1)]/l$ will have a minimum corresponding to an
equilibrium chain length. In two dimensions these chains will unify in
square platelets. Thus, our model contains the mechanism for the size
calibration of atomic clusters which was phenomenologically described
in Ref.\ \cite{lannoo}. In 2D this conclusion as well as the formula of
Ref.\ \cite{lannoo} were verified with the use of the Monte Carlo
simulation \cite{molphys}.

Because the entropic contribution $S_l$ (see Fig. \ref{fig1}a) is
practically linear in $l$ according to Eq.\ (\ref{V_l}) which has the
form of the discrete second derivative it essentially contributes only
into the pair interaction $V_2$ (the multiatom contributions were found
to be $\sim10\%$). Thus, the relaxation entropy formally amounts to
effective NN interatomic repulsion which grows linearly with
temperature. The entropic forces of this kind were earlier discovered
in alloys (see Appendix E of Ref.\ \cite{rmp2} and references to
earlier literature therein).

At low temperature the main contributions into $V_l$ for $l\geq3$ come
from $W_l$ (Fig.\ref{fig1}b). Because the r. h. s.  of Eq.\ (\ref{V_l})
has the form of the discrete second derivative and the curve $W_l(l)$
is concave, all multiatom interactions are repulsive. On the other
hand, in the case under consideration when atoms are supposed to
assemble into clusters, there should exist attractive interactions
which in our model are necessarily the pair ones [see Eq.\ (\ref{U})].
\section{Examples from metallic heteroepitaxial systems}
To illustrate the above formalism with realistic examples of strained
epitaxy we consider two metallic heteroepitaxial systems---Ag/Pt and
Pt/Co---which currently are being actively studied both experimentally
\cite{AgPt,1D2Dexp,Pt997} and theoretically \cite{CoPtth,PtCoth2}.  For
simplicity we considered 1D case and used the geometry of Ref.
\cite{1D2Dth} where the growth on the steps of the closed packed (111)
vicinal surface was studied. The position of a deposited atom was
relaxed to its equilibrium value in order to find the value $k_s$ as
the second derivative of the potential near equilibrium. The many-body
``potentials'' and corresponding parameters were taken from
Ref.\ \cite{1D2Dth} for the Ag/Pt system and from Ref.\ \cite{PtCoth2}
for the Pt/Co system. These potentials were devised specifically to
application in the heteroepitaxy and for the (111) surfaces so we
expect our results below are quite reliable. Other parameters listed in
Table \ref{table1} were calculated for the atomic pairs relaxed only in
vertical position because in our approach we need the first and second
derivatives [see discussion after Eq.\ (\ref{U})] calculated in the
center of of symmetry of the 1D cell. In both calculations only the NN
interactions were taken into account because the NNN ones were found to
be negligeable.

\begin{figure}
\includegraphics[viewport = 195 520 415 740, scale = 1.]{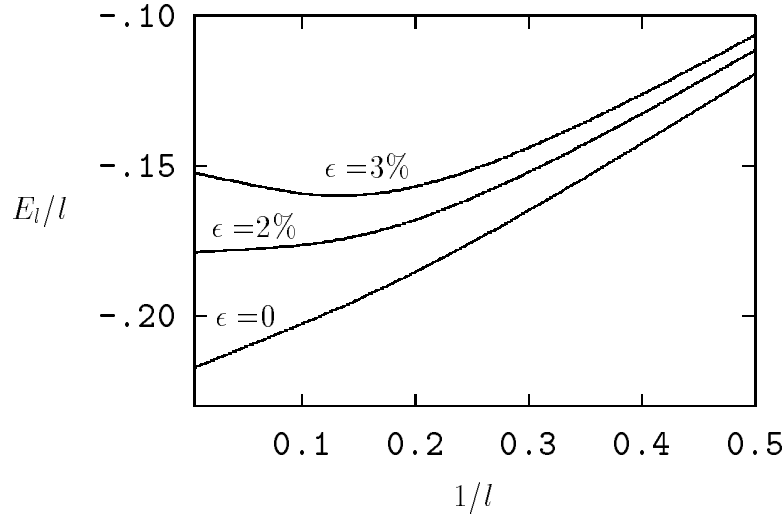} 
\caption{Length dependence of the reduced energy at zero temperature
of the Pt monatomic
chains for different values of compression $\epsilon$ of the Co substrate.}
\label{fig2}
\end{figure}

According to our calculations the Ag/Pt system has the following
parameters: (the energy unit is eV, the length unit \AA):
$V^p_{\text{NN}} \approx -0.57$, $V^p_{\text{NNN}} \approx
-8.8\cdot10^{-3} $, $k_pf^2 \approx 0.72$, and $\alpha\approx 3.7$. The
large value of $\alpha$ means that the relaxation of the strain is very
weak (see Fig.\ \ref{fig1}) so there is no size calibration with the
above parameters.  The crucial parameter $\alpha$, however, can
strongly vary in different systems.  For example, according to our
estimates based on the potentials of Ref.\ \cite{PtCoth2}, in Pt/Co
$\alpha\simeq0.32$ is an order of magnitude smaller.  Because of this
the system is quite close to the self-assembly but the misfit strain is
still too small. To enhance the misfit, in our model calculations we
assumed that the Co underlayer is further compressed (e. g., by means
of deposition on an appropriate substrate) by the factor $1-\epsilon$.
(see Table \ref{table1} and Fig.\ \ref{fig2}). But for $\epsilon$ below
the critical value $\epsilon_c\approx2\%$ there is no self-assembly at
$T=0$ because there is no minimum in $E_l(T=0)/l$ (see
Ref.\ \cite{lannoo}).  However, because of the $T$-dependent entropic
contribution in the effective energy (\ref{F2}), slightly below
$\epsilon_c$ the system does exhibit the self-assembly at intermediate
temperatures which, however, disappears as $T\rightarrow0$ (see
Fig.\ \ref{fig2}). Thus, our model predicts a new phenomenon which may
be called the entropy driven transient self-assembly. For higher values
of strain the $E_l(T=0)/l$ curve does have a minimum (see
Fig.\ \ref{fig2}) in which case the qualitative analysis of
Ref.\ \cite{lannoo} fully applies.
\begin{table}
\label{table1}
\caption{Parameters corresponding to the Pt/Co heteroepitaxial
system }
\begin{center}
\begin{tabular*}{8cm}{@{\extracolsep{\fill}}cccc}
$\epsilon$&$\alpha$&$k_pf^2$ (eV)&$V_{NN}$ (eV)\\
\hline
 0\%& 0.32&0.049&-0.218\\
 2\%& 0.25&0.100&-0.179\\
 3\%&0.22& 0.136&-0.152\\
\hline
\end{tabular*}
\end{center}
\end{table}
\section{conclusion}
In this paper we considered a simple model of strained 1D epitaxy and
have shown that similarly to the 2D case the mechanism of self-assembly
and size calibration proposed in Ref.\ \cite{lannoo} is operative also
in this case. We considered only two explicit examples for which there
exist in literature reliable interatomic potentials and in one system
already found the size calibration for stressed substrate. In our
opinion, this shows that the above phenomena should be as common in 1D
as they are in 2D heteroepitaxy.  Further argument in favor of this
conclusion is that for the size calibration to be plausible the crucial
parameter $\alpha=k_s/k_p$ should be as small as possible. This favors
small values of $k_s$. However, the geometry considered in the present
paper is not quite favorable because of the high coordination of the
atoms deposited at the steps with 5 NN atoms of the substrate (see
Ref.\ \cite{1D2Dth}) which enhances $k_s$. It may be hoped that with
lower coordination of the deposited atoms (as in the case of 1D
structures of Ref.\ \cite{nat_chains}) the conditions for the size
calibration will be more favorable.
\begin{acknowledgments}
One of us (VT) expresses his gratitude to University Louis Pasteur de
Strasbourg and IPCMS for their hospitality.
\end{acknowledgments}
\appendix
\section{}
\label{appendix}
According to Eq.\ (\ref{kD})
\begin{equation} 
\label{ D}
D_l=\left(\underbrace{
\begin{array}{ccccccc} 
 1+\alpha &   -1   &  0   &\dots&0&0&0\\
   -1    & 2+\alpha&  -1   &0&\dots&0&0\\
   0    &   -1   &  2+\alpha&-1&0&\dots&0\\
\vdots &\vdots  &\ddots &  \ddots &\ddots&\vdots&\vdots\\
0& \dots &0&  -1  & 2+\alpha &   -1    &   0   \\
0&0&\dots   &0     & -1   & 2+\alpha &   -1 \\
0&0&0&\dots  &      0   &   -1    & 1+\alpha
\end{array}}_{\textstyle l}
\right),
\end{equation} 
where $\alpha=k_s/k_p$.  Because matrices $D_l$ are tri-diagonal, their
determinants satisfy recurrence relations which can be used to
calculate all quantities entering Eqs.\ (\ref{Frelax}) and
(\ref{V_l}).  The diagonal element of the matrix $G_l=1/D_l$ is
\begin{equation} 
\label{ G0}
G^{(0)}_{l}=r_{l-1}/\det D_l,
\end{equation} 
where $r_{l-1}$ is the determinant of the matrix obtained from $D_l$ 
by crossing out its first row and the first column:
\begin{equation} 
\label{ m}
r_{l-1}=\left|\underbrace{
\begin{array}{cccccc} 
 2+\alpha &   -1   &  0   &\dots&0&0\\
   -1    & 2+\alpha&  -1   &0&\dots&0\\
   0    &   -1   &  2+\alpha&-1&0&0\\
\vdots &\vdots  &\ddots &  \ddots &\ddots&\vdots\\
0& \dots &0&  -1  & 2+\alpha &   -1 \\
0&0&\dots   &0     & -1   & 1+\alpha\\
\end{array}}_{\textstyle l-1}
\right|.
\end{equation} 
Expanding $\det D_l$ with respect to the first row we get
\begin{equation} 
\label{ detD}
\det D_l = (1+\alpha)r_{l-1}-r_{l-2}.
\end{equation} 
Comparing this with Eq.\ (\ref{ G0}) we get
\begin{equation} 
G^{(0)}_{l}=1/(1+\alpha-d_{l-1}),
\end{equation} 
where $d_{l-1}=r_{l-2}/r_{l-1}$.
Now, expanding determinant Eq.\ (\ref{ m}) with respect to the elements
of the first row we get the following three term recurrence relation
\begin{equation} 
\label{ mm}
r_{l-1}=(2+\alpha)r_{l-2}-r_{l-3}.
\end{equation} 
But because Eq.\ (\ref{ G0}) includes only the ratio $r_{l-2}/r_{l-1}$
by dividing Eq.\ (\ref{ mm}) by $r_{l-2}$ we can transform it to a
simpler form
\begin{equation} 
\frac{1}{d_{l-1}}=2+\alpha - d_{l-2}.
\end{equation} 
The off-diagonal matrix element $G_l^{(l-1)}$ of the inverse matrix
$G_l$ is equal to the ratio of the determinant of the matrix obtained
from $D_l$ by crossing out its first row and the last column multiplied
by $(-1)^{l+1}$ and divided by $\det D_l$. As is easy to see from Eq.\
(\ref{ D}) the matrix thus obtained is a triangular matrix with the
diagonal elements being all equal to -1. Hence its determinant is equal
to $(-1)^{l-1}$ and
\begin{equation} 
G_l^{(l-1)}=1/\det D_l = \det G_l.
\end{equation} 
From Eq.\ (\ref{ detD}) we get
\begin{equation} 
G_l^{(l-1)}=\frac{1}{\det D_l}=\frac{1}{r_{l-1}(1+\alpha-d_{l-1})}
=\frac{b_{l-1}}{1+\alpha-d_{l-1}}=b_{l-1}G^{(0)}_{l},
\end{equation} 
where $b_{l-1}=1/r_{l-1}=d_{l-1}b_{l-2}$ (see the definition of
$d_{l-1}$ above).
Finally, making the shift $l-1\to l+1$ the above formulas 
can be summarized as
\begin{equation} 
\label{ GG}
\left\{
\begin{array}{lcl}
G^{(0)}_{l+1} &=& 1/(1+\alpha-d_l)\nonumber\\
G^{(l)}_{l+1} &=& \det G_{l+1} =G^{(0)}_{l+1}b_l,\nonumber
\end{array}
\right.
\end{equation} 
where $d_l$ and $b_l$ are generated by the recursion 
relations
\begin{equation} 
\label{ recursion}
d_{l+1} = 1/(2+\alpha - d_l)\quad\mbox{and}\quad  
b_{l+1} = d_{l+1}b_l
\end{equation} 
initialized by $d_0=b_0=1$.


\begin{thebibliography}{18}
\expandafter\ifx\csname natexlab\endcsname\relax\def\natexlab#1{#1}\fi
\expandafter\ifx\csname bibnamefont\endcsname\relax
  \def\bibnamefont#1{#1}\fi
\expandafter\ifx\csname bibfnamefont\endcsname\relax
  \def\bibfnamefont#1{#1}\fi
\expandafter\ifx\csname citenamefont\endcsname\relax
  \def\citenamefont#1{#1}\fi
\expandafter\ifx\csname url\endcsname\relax
  \def\url#1{\texttt{#1}}\fi
\expandafter\ifx\csname urlprefix\endcsname\relax\def\urlprefix{URL }\fi
\providecommand{\bibinfo}[2]{#2}
\providecommand{\eprint}[2][]{\url{#2}}

\bibitem[{\citenamefont{Eaglesham and Cerullo}(1990)}]{discovery}
\bibinfo{author}{\bibfnamefont{D.~J.} \bibnamefont{Eaglesham}}
  \bibnamefont{and} \bibinfo{author}{\bibfnamefont{M.}~\bibnamefont{Cerullo}},
  \bibinfo{journal}{Phys. Rev. Lett.} \textbf{\bibinfo{volume}{64}},
  \bibinfo{pages}{1943} (\bibinfo{year}{1990});
\bibinfo{author}{\bibfnamefont{Y.-W.} \bibnamefont{Mo}},
  \bibinfo{author}{\bibfnamefont{D.~E.} \bibnamefont{Savage}},
  \bibinfo{author}{\bibfnamefont{B.~S.} \bibnamefont{Swartzentruber}},
  \bibnamefont{and} \bibinfo{author}{\bibfnamefont{M.~G.}
  \bibnamefont{Lagally}}, \bibinfo{journal}{Phys. Rev. Lett.}
  \textbf{\bibinfo{volume}{65}}, \bibinfo{pages}{1020} (\bibinfo{year}{1990});
\bibinfo{author}{\bibfnamefont{S.}~\bibnamefont{Guha}},
  \bibinfo{author}{\bibfnamefont{A.}~\bibnamefont{Madhukar}}, \bibnamefont{and}
  \bibinfo{author}{\bibfnamefont{K.~C.} \bibnamefont{Rajkumar}},
  \bibinfo{journal}{Appl. Phys. Lett.} \textbf{\bibinfo{volume}{57}},
  \bibinfo{pages}{2110} (\bibinfo{year}{1990}).

\bibitem[{\citenamefont{Notzel et~al.}(1994)\citenamefont{Notzel, Temmyo, and
  Tamamura}}]{dots}
\bibinfo{author}{\bibfnamefont{R.}~\bibnamefont{Notzel}},
  \bibinfo{author}{\bibfnamefont{J.}~\bibnamefont{Temmyo}}, \bibnamefont{and}
  \bibinfo{author}{\bibfnamefont{T.}~\bibnamefont{Tamamura}},
  \bibinfo{journal}{Nature (London)} \textbf{\bibinfo{volume}{369}},
  \bibinfo{pages}{131} (\bibinfo{year}{1994});
\bibinfo{author}{\bibfnamefont{D.}~\bibnamefont{Leonard}
  \bibnamefont{et~al.}},
  \bibinfo{journal}{Appl. Phys. Lett.} \textbf{\bibinfo{volume}{63}},
  \bibinfo{pages}{3203} (\bibinfo{year}{1993}).

\bibitem[{\citenamefont{Orlov et~al.}(1996)\citenamefont{Orlov, Amlani, Lent,
  and Snider}}]{devices}
\bibinfo{author}{\bibfnamefont{A.~O.} \bibnamefont{Orlov}},
  \bibinfo{author}{\bibfnamefont{I.}~\bibnamefont{Amlani}},
  \bibinfo{author}{\bibfnamefont{C.~S.} \bibnamefont{Lent}},
  \bibnamefont{and} \bibinfo{author}{\bibfnamefont{G.~L.}
  \bibnamefont{Snider}}, \bibinfo{journal}{Science}
  \textbf{\bibinfo{volume}{277}}, \bibinfo{pages}{928} (\bibinfo{year}{1997}).

\bibitem[{\citenamefont{van~der Merwe et~al.}(1994)\citenamefont{van~der Merwe,
  T\"onsing, and Stoop}}]{mismatch}
\bibinfo{author}{\bibfnamefont{J.~H.} \bibnamefont{van~der Merwe}},
  \bibinfo{author}{\bibfnamefont{D.~L.} \bibnamefont{T\"onsing}},
  \bibnamefont{and} \bibinfo{author}{\bibfnamefont{P.~M.} \bibnamefont{Stoop}},
  \bibinfo{journal}{Surf. Sci.} \textbf{\bibinfo{volume}{312}},
  \bibinfo{pages}{387} (\bibinfo{year}{1994});
\bibinfo{author}{\bibfnamefont{M.}~\bibnamefont{Henzler}},
  \bibinfo{journal}{Surf. Sci.} \textbf{\bibinfo{volume}{357--358}},
  \bibinfo{pages}{809} (\bibinfo{year}{1996});
\bibinfo{author}{\bibfnamefont{S.}~\bibnamefont{Tan}},
  \bibinfo{author}{\bibfnamefont{A.}~\bibnamefont{Ghazali}}, \bibnamefont{and}
  \bibinfo{author}{\bibfnamefont{J.~C.~S.} \bibnamefont{L\'evy}},
  \bibinfo{journal}{Surf. Sci.} \textbf{\bibinfo{volume}{369}},
  \bibinfo{pages}{360} (\bibinfo{year}{1996}).

\bibitem[{\citenamefont{Priester and Lannoo}(1995)}]{lannoo}
\bibinfo{author}{\bibfnamefont{C.}~\bibnamefont{Priester}} \bibnamefont{and}
  \bibinfo{author}{\bibfnamefont{M.}~\bibnamefont{Lannoo}},
  \bibinfo{journal}{Phys. Rev. Lett.} \textbf{\bibinfo{volume}{75}},
  \bibinfo{pages}{93} (\bibinfo{year}{1995}).

\bibitem[{\citenamefont{Khor and Sarma}(2000)}]{2D3D}
\bibinfo{author}{\bibfnamefont{K.~E.} \bibnamefont{Khor}} \bibnamefont{and}
  \bibinfo{author}{\bibfnamefont{S.~D.} \bibnamefont{Sarma}},
  \bibinfo{journal}{Phys. Rev. B} \textbf{\bibinfo{volume}{62}},
  \bibinfo{pages}{16657} (\bibinfo{year}{2000}).

\bibitem[{\citenamefont{Segovia et~al.}(1999)\citenamefont{Segovia, Purdie,
  Hensenberger, and Baer}}]{nat_chains}
\bibinfo{author}{\bibfnamefont{P.}~\bibnamefont{Segovia}},
  \bibinfo{author}{\bibfnamefont{D.}~\bibnamefont{Purdie}},
  \bibinfo{author}{\bibfnamefont{M.}~\bibnamefont{Hensenberger}},
  \bibnamefont{and} \bibinfo{author}{\bibfnamefont{Y.}~\bibnamefont{Baer}},
  \bibinfo{journal}{Nature (London)} \textbf{\bibinfo{volume}{402}},
  \bibinfo{pages}{504} (\bibinfo{year}{1999}).

\bibitem[{\citenamefont{Dorantes-D{\'a}vila and Pastor}(1998)}]{magnetism}
\bibinfo{author}{\bibfnamefont{J.}~\bibnamefont{Dorantes-D{\'a}vila}}
  \bibnamefont{and} \bibinfo{author}{\bibfnamefont{G.~M.}
  \bibnamefont{Pastor}}, \bibinfo{journal}{Phys. Rev. Lett.}
  \textbf{\bibinfo{volume}{81}}, \bibinfo{pages}{208} (\bibinfo{year}{1998}).

\bibitem[{\citenamefont{Snyman and van~der Merwe}(1974)}]{2d1d}
\bibinfo{author}{\bibfnamefont{J.~A.} \bibnamefont{Snyman}} \bibnamefont{and}
  \bibinfo{author}{\bibfnamefont{J.~H.} \bibnamefont{van~der Merwe}},
  \bibinfo{journal}{Surf. Sci.} \textbf{\bibinfo{volume}{45}},
  \bibinfo{pages}{619} (\bibinfo{year}{1974}).

\bibitem[{\citenamefont{Korutcheva et~al.}(2000)\citenamefont{Korutcheva,
  Turiel, and Markov}}]{model2}
\bibinfo{author}{\bibfnamefont{E.}~\bibnamefont{Korutcheva}},
  \bibinfo{author}{\bibfnamefont{A.~M.} \bibnamefont{Turiel}},
  \bibnamefont{and} \bibinfo{author}{\bibfnamefont{I.}~\bibnamefont{Markov}},
  \bibinfo{journal}{Phys. Rev. B} \textbf{\bibinfo{volume}{61}},
  \bibinfo{pages}{16890} (\bibinfo{year}{2000}).

\bibitem[{\citenamefont{Erwin et~al.}(1999)\citenamefont{Erwin, Baski, Whitman,
  and Rudd}}]{baskiPRL}
\bibinfo{author}{\bibfnamefont{S.~C.} \bibnamefont{Erwin}},
  \bibinfo{author}{\bibfnamefont{A.~A.} \bibnamefont{Baski}},
  \bibinfo{author}{\bibfnamefont{L.~J.} \bibnamefont{Whitman}},
  \bibnamefont{and} \bibinfo{author}{\bibfnamefont{R.~E.} \bibnamefont{Rudd}},
  \bibinfo{journal}{Phys. Rev. Lett.} \textbf{\bibinfo{volume}{83}},
  \bibinfo{pages}{1818} (\bibinfo{year}{1999}).

\bibitem[{\citenamefont{Uebing and Homer}(1991)}]{thereference}
\bibinfo{author}{\bibfnamefont{C.}~\bibnamefont{Uebing}} \bibnamefont{and}
  \bibinfo{author}{\bibfnamefont{R.}~\bibnamefont{Homer}}, \bibinfo{journal}{J.
  Chem. Phys.} \textbf{\bibinfo{volume}{95}}, \bibinfo{pages}{7626}
  (\bibinfo{year}{1991}).

\bibitem[{\citenamefont{Orr et~al.}(1992)\citenamefont{Orr, Kessler, Snyder,
  and Sander}}]{model3}
\bibinfo{author}{\bibfnamefont{B.~G.} \bibnamefont{Orr}},
  \bibinfo{author}{\bibfnamefont{D.}~\bibnamefont{Kessler}},
  \bibinfo{author}{\bibfnamefont{C.}~\bibnamefont{Snyder}}, \bibnamefont{and}
  \bibinfo{author}{\bibfnamefont{L.}~\bibnamefont{Sander}},
  \bibinfo{journal}{Europhys. Lett.} \textbf{\bibinfo{volume}{19}},
  \bibinfo{pages}{33} (\bibinfo{year}{1992}).

\bibitem[{\citenamefont{Ratsch and Zangwill}(1993)}]{model1}
\bibinfo{author}{\bibfnamefont{C.}~\bibnamefont{Ratsch}} \bibnamefont{and}
  \bibinfo{author}{\bibfnamefont{A.}~\bibnamefont{Zangwill}},
  \bibinfo{journal}{Surf. Sci.} \textbf{\bibinfo{volume}{293}},
  \bibinfo{pages}{123} (\bibinfo{year}{1993}).

\bibitem[{\citenamefont{Venables}(1987)}]{einstein1}
\bibinfo{author}{\bibfnamefont{J.~A.} \bibnamefont{Venables}},
  \bibinfo{journal}{Phys. Rev. B} \textbf{\bibinfo{volume}{36}},
  \bibinfo{pages}{4153} (\bibinfo{year}{1987}).

\bibitem[{\citenamefont{Doye and Calvo}(2000)}]{einstein2}
\bibinfo{author}{\bibfnamefont{J.~P.~K.} \bibnamefont{Doye}} \bibnamefont{and}
  \bibinfo{author}{\bibfnamefont{F.}~\bibnamefont{Calvo}},
  \bibinfo{journal}{Phys. Rev. Lett.} \textbf{\bibinfo{volume}{86}},
  \bibinfo{pages}{3570} (\bibinfo{year}{2001}).

\bibitem[{\citenamefont{Tokar and Dreyss\'e}(2002)}]{molphys}
\bibinfo{author}{\bibfnamefont{V.~I.} \bibnamefont{Tokar}} \bibnamefont{and}
  \bibinfo{author}{\bibfnamefont{H.}~\bibnamefont{Dreyss\'e}},
  \bibinfo{journal}{Molec. Phys.} \textbf{\bibinfo{volume}{100}},
  \bibinfo{pages}{3151} (\bibinfo{year}{2002}).

\bibitem[{\citenamefont{van~de Walle and Ceder}(2002)}]{rmp2}
\bibinfo{author}{\bibfnamefont{A.}~\bibnamefont{van~de Walle}}
  \bibnamefont{and} \bibinfo{author}{\bibfnamefont{G.}~\bibnamefont{Ceder}},
  \bibinfo{journal}{Rev. Mod. Phys} \textbf{\bibinfo{volume}{74}},
  \bibinfo{pages}{11} (\bibinfo{year}{2002}).

\bibitem[{\citenamefont{Becker et~al.}(1993)\citenamefont{Becker, Rosenfeld,
  Poelsema, and Comsa}}]{AgPt}
\bibinfo{author}{\bibfnamefont{A.~F.} \bibnamefont{Becker}},
  \bibinfo{author}{\bibfnamefont{G.}~\bibnamefont{Rosenfeld}},
  \bibinfo{author}{\bibfnamefont{B.}~\bibnamefont{Poelsema}}, \bibnamefont{and}
  \bibinfo{author}{\bibfnamefont{G.}~\bibnamefont{Comsa}},
  \bibinfo{journal}{Phys. Rev. Lett.} \textbf{\bibinfo{volume}{70}},
  \bibinfo{pages}{477} (\bibinfo{year}{1993}).

\bibitem[{\citenamefont{Gambardella
  et~al.}(2000{\natexlab{a}})\citenamefont{Gambardella, Blanc, Brune, Kuhnke,
  and Kern}}]{1D2Dexp}
\bibinfo{author}{\bibfnamefont{P.}~\bibnamefont{Gambardella}},
  \bibinfo{author}{\bibfnamefont{M.}~\bibnamefont{Blanc}},
  \bibinfo{author}{\bibfnamefont{H.}~\bibnamefont{Brune}},
  \bibinfo{author}{\bibfnamefont{K.}~\bibnamefont{Kuhnke}}, \bibnamefont{and}
  \bibinfo{author}{\bibfnamefont{K.}~\bibnamefont{Kern}},
  \bibinfo{journal}{Phys. Rev. B} \textbf{\bibinfo{volume}{61}},
  \bibinfo{pages}{2254} (\bibinfo{year}{2000}{\natexlab{a}}).

\bibitem[{\citenamefont{Gambardella
  et~al.}(2000{\natexlab{b}})\citenamefont{Gambardella, Blanc, B{\"u}rgi,
  Kuhnke, and Kern}}]{Pt997}
\bibinfo{author}{\bibfnamefont{P.}~\bibnamefont{Gambardella}},
  \bibinfo{author}{\bibfnamefont{M.}~\bibnamefont{Blanc}},
  \bibinfo{author}{\bibfnamefont{L.}~\bibnamefont{B{\"u}rgi}},
  \bibinfo{author}{\bibfnamefont{K.}~\bibnamefont{Kuhnke}}, \bibnamefont{and}
  \bibinfo{author}{\bibfnamefont{K.}~\bibnamefont{Kern}},
  \bibinfo{journal}{Surf. Sci.} \textbf{\bibinfo{volume}{449}},
  \bibinfo{pages}{93} (\bibinfo{year}{2000}{\natexlab{b}}).

\bibitem[{\citenamefont{Goyhenex and Tr{\'e}glia}(2000)}]{CoPtth}
\bibinfo{author}{\bibfnamefont{C.}~\bibnamefont{Goyhenex}} \bibnamefont{and}
  \bibinfo{author}{\bibfnamefont{G.}~\bibnamefont{Tr{\'e}glia}},
  \bibinfo{journal}{Surf. Sci.} \textbf{\bibinfo{volume}{446}},
  \bibinfo{pages}{272} (\bibinfo{year}{2000}).

\bibitem[{\citenamefont{Goyhenex et~al.}(1999)\citenamefont{Goyhenex, Bulou,
  Deville, and Tr{\'e}glia}}]{PtCoth2}
\bibinfo{author}{\bibfnamefont{C.}~\bibnamefont{Goyhenex}},
  \bibinfo{author}{\bibfnamefont{H.}~\bibnamefont{Bulou}},
  \bibinfo{author}{\bibfnamefont{J.-P.} \bibnamefont{Deville}},
  \bibnamefont{and}
  \bibinfo{author}{\bibfnamefont{G.}~\bibnamefont{Tr{\'e}glia}},
  \bibinfo{journal}{Phys. Rev. B} \textbf{\bibinfo{volume}{60}},
  \bibinfo{pages}{2781} (\bibinfo{year}{1999}).

\bibitem[{\citenamefont{Picaud et~al.}(2000)\citenamefont{Picaud, Ramseyer,
  Girardet, and Jensen}}]{1D2Dth}
\bibinfo{author}{\bibfnamefont{F.}~\bibnamefont{Picaud}},
  \bibinfo{author}{\bibfnamefont{C.}~\bibnamefont{Ramseyer}},
  \bibinfo{author}{\bibfnamefont{C.}~\bibnamefont{Girardet}}, \bibnamefont{and}
  \bibinfo{author}{\bibfnamefont{P.}~\bibnamefont{Jensen}},
  \bibinfo{journal}{Phys. Rev. B} \textbf{\bibinfo{volume}{61}},
  \bibinfo{pages}{16154} (\bibinfo{year}{2000}).

\end{thebibliography}
\end{document}